\documentclass[aps,prl,twocolumn,superscriptaddress,showpacs]{revtex4}
\usepackage{placeins}
\usepackage{graphicx}
\usepackage{amsmath}
\usepackage{amssymb}
\usepackage{times}
\usepackage{color}

\begin{document}

\preprint{}

\title{Dynamic system of biological nitrogen fixation in a strange-attractor regime}

\author{V.P.~Gachok}
\email{aszhokhin@bitp.kiev.ua}
\affiliation{Bogolyubov Institute for Theoretical Physics of the National Academy of Sciences of Ukraine}%

\author{A.S.~Zhokhin}
\email{aszhokhin@bitp.kiev.ua} 
\affiliation{Bogolyubov Institute for Theoretical Physics of the National Academy of Sciences of Ukraine}%

\begin{abstract}
In this paper, we investigate a mathematical modeling of biochemical processes which makes it possible to determine the adaptation of a biosystem to open environmental conditions.

\end{abstract}


\maketitle

A mathematical modeling of biochemical processes makes it possible to determine the adaptation of a biosystem to open environmental conditions. The mass-balance equations incorporate the nonlinear cooperative nature of the transformation processes. As a result, we have a self-regulation problem for dynamic systems far from equilibrium. The phase-portrait method, in particular the Poincare mapping, is effective for studying such nonlinear problems. In the present we wish to propose a model of this type, which is defined by the following system of nonlinear differential equations:

\begin{equation}
\label{eq1} \partial_t S=\frac{S_0 K}{K+S+\Psi ^4 }-l_1 V(S)V(S_6) -\alpha _S S
\end{equation}\vspace*{-7mm}
\begin{equation}
\label{eq2} \partial_t S_{1} =l_1 V(S)V(S_6) -l_2 V(N^2)V(S_1^2)
\end{equation}\vspace*{-7mm}
\begin{equation}
\label{eq3} \partial_t S_{2} =l_2 V(N^2)V(S_1^2 ) -l_3 V(N)V(S_2)
\end{equation}\vspace*{-7mm}
\begin{equation}
\label{eq4} \partial_t S_{3} =l_3 V(N)V(S_2) -4l_4 V(L_1-T)V(S_3)
\end{equation}\vspace*{-7mm}
\begin{equation}
\label{eq5} \partial_t S_{4} =4l_4 V(L_1-T)V(S_3)-l_5 V(N) \frac{S_4}{1+S_4^2 +MS_6 }
\end{equation}\vspace*{-7mm}
\begin{equation}
\label{eq6} \partial_t S_{5} =l_5 V(N) \frac{S_4}{1+S_4^2 +MS_6 } -l_6 V(N)V(S_5)
\end{equation}\vspace*{-7mm}
\begin{equation}
\label{eq7} \partial_t S_{6} =l_6 V(N)V(S_5) -l_1 V(S)V(S_6)
\end{equation}\vspace*{-7mm}
\begin{eqnarray}
\label{eq8} \partial_t N =k_5 V(E_1)V(L_2 -N)-l_2 V(N^2)V(S_1^2) \nonumber \\
-k_3 V(T^4)\frac{E_3}{1+E_3 +O_2} -l_5 V(N) \frac{S_4}{1+S_4^2 +MS_6 }\nonumber \\
 -l_6 V(N)V(S_5) +l_7 V(Q)V(L_2 -N)
\end{eqnarray}\vspace*{-7mm}
\begin{equation}
\label{eq9} \partial_t Q =6k_2 \frac{V(O_2)V(L_1-Q)}{1+\Psi^2} -l_7 V(Q)V(L_2 -N)
\end{equation}\vspace*{-7mm}
\begin{eqnarray}
\label{eq10}
\partial_t T=k_4 V(N_2)V(\Psi^4 )\frac{E_4}{1+E_4+H_2/2} \nonumber \\
+l_8 V(L_1-T)V(\Psi^4)-4k_3 V(T^4)\frac{E_3}{1+E_3 +O_2} \nonumber \\
-2k_6 \frac{E_3}{E_3 +K_{11}}V(T^2)V(\Psi^2 )
\end{eqnarray}\vspace*{-7mm}
\begin{eqnarray}
\label{eq11}
\partial_t \Psi =4l_7 V(Q)V(L_2 -N) -4l_8 V(L_1-T)V(\Psi^4) \nonumber \\
 -k_1 V(E_1)V(\Psi) -4k_4 V(N_2)V(\Psi^4 )\frac{E_4}{1+E_4+H_2/2}\nonumber \\ -2k_6 \frac{E_3}{E_3 +K_{11}}V(T^2)V(\Psi^2) -\alpha_{\Psi} \Psi 
\end{eqnarray}\vspace*{-7mm}

\begin{eqnarray}
\label{eq12}
\partial_t O_2= O_{20}\frac{K_5}{K_5+O_2}-k_2 \frac{V(O_2)V(L_1-Q)}{1+\Psi^2}
\end{eqnarray}\vspace*{-7mm}

\begin{eqnarray}
\label{eq13}
\partial_t E_{1} =E_{10}\frac{A}{A+\beta_1}\frac{K_1}{K_1+A^4}-k_1 V(E_1)V(\Psi)\hspace*{7mm} \nonumber \\ +k_4 V(N_2)V(\Psi^4 )\frac{E_4}{1+E_4+H_2/2} -k_5 V(E_1)V(L_2 -N) \nonumber \hspace*{4mm} \\
+k_6 \frac{E_3}{E_3 +K_{11}}V(T^2)V(\Psi^2 )-k_7 V(E_1)V(H_2) -\alpha_{E_1} E_1 \hspace*{4mm}
\end{eqnarray}\vspace*{-7mm}

\begin{eqnarray}
\label{eq14}
\partial_t E_3 =k_1 V(E_1)V(\Psi) -k_3 V(T^4)\frac{E_3}{1+E_3 +O_2} \nonumber\\
+k_5 V(E_1)V(L_2 -N) -k_6 \frac{E_3}{E_3 +K_{11}}V(T^2)V(\Psi^2 ) \nonumber\\
+k_7 V(E_1)V(H_2) 
\end{eqnarray}\vspace*{-7mm}
\begin{eqnarray}
\label{eq15}
\partial_t E_4=k_3 V(T^4)\frac{E_3}{1+E_3 +O_2} \nonumber\\
-k_4 V(N_2)V(\Psi^4)\frac{E_4}{1+E_4+H_2/2} 
\end{eqnarray}\vspace*{-7mm}
\begin{eqnarray}
\label{eq16}
\partial_t N_2 =N_{20}\frac{K_2} {K_2+N_2}-\nonumber\\
k_4V(N_2)V(\Psi^4)\frac{E_4}{1+E_4+H_2/2} -\alpha _{N_2} N_2 
\end{eqnarray}\vspace*{-7mm}
\begin{eqnarray}
\label{eq17}
\partial_t A=k_4 V(N_2)V(\Psi^4 )\frac{E_4}{1+E_4+H_2/2} -\alpha_A A
\end{eqnarray}\vspace*{-7mm}
\begin{eqnarray}
\label{eq18}
\partial_t H_2 =k_6 \frac{E_3}{E_3 +K_{11}}V(T^2)V(\Psi^2 )\nonumber\\
 -k_7 V(E_1)V(H_2)-\alpha _{H_{2} } H_{2} 
\end{eqnarray}\vspace*{-7mm}

\vspace*{3mm}

\noindent where \textit{$V(x)=x/(x+1)$}. The dynamic variables  $S(t),...,S_6(t)$ describe the Krebs cycle. The variables  $Q(t),\Psi (t),T(t),O_{2}(t)$ describe the respiratory chain, the kinetic membrane potential, the ATP concentration, and the concentration of molecular oxygen, respectively ~\cite{gachok81,gach_zhokh85}.

This block of the model is responsible for the describing the bioenergetics of a cell. The crucial enzyme nitrogenase participates in its three forms ~\cite{pnfx82,shilov82,likht78} $E_1$, $E_3$, $E_4$.
Its activity is regulated by molecular hydrogen $H_2$ and molecular nitrogen $N_2$, while the biosynthesis is regulated by ammonia A, which is present in the biosystem.

Serving as an electron donor here is $NAD \cdot H_2 \equiv N$.

The optimal regime of the biosystem is described by the following set of parameter values:

\begin{eqnarray*}
\label{eq19}
l_1 =0.3,\ l_{2} =0.3,\ l_{3} =0.3,\\
l_4 =0.03,\ l_{5} =0.3,\ l_{6} =0.3,\\
l_7 =0.4,\ l_8 =0.2,\ k_1 =0.01,\\
k_2 =0.05,\ k_3 =0.05,\ k_4 =0.06,\\
k_5 =0.05,\ k_6 =0.03,\ k_7 =0.003,\\
L_1 =3,\ L_2 =2,\ L_3 =2,\\
K=0.3,\ K_1=0.006,\ K_2=0.3,\\
\alpha_S=0.005,\ \alpha_A=0.0045,\ O_{20}=0.02,\\
E_{10} =0.02,\ \beta =0.1,\ N_{20} =0,2.\\
\end{eqnarray*}\nonumber\vspace*{-7mm}

The kinetics of the biosystem exhibits a self-regulation. Under the open environmental conditions conditions described by the parameters $S_0 ,O_{20} ,N_{20} ,\alpha_{S},\alpha_1,\alpha_2,...$ the phase-portrait of the biosystem includes a structural adapter characterized by stable oscillations and does not dependent from a initial state of the biosystem (Cauchy data).

The presence of oscillations in the biosystem is determined primarily by the interaction of self-regulated subsystems: the Krebs cycle, the respiratory chain, and the nitrogenase complex.

Furthermore, the process of biosynthesis of nitrogenase in form of $E_1$ depends on the ammonia concentration in a non-monotonic way. This circumstance is expressed by the nonlinear term in \eqref{eq13}:
\vspace*{-2mm} $$E_{10}\frac{A}{A+\beta_1}\frac{K_1}{K_1+A^4}$$ 

At small values of A, the synthesis of $E_{1} $is induced, while a large value of $A$ results  in a repression of $E_{1} $.

At the pumping parameter value$S_{0} =0.07$, a strange attractor exists in the phase space of the system. Figure 1 shows a projection of the phase portrait of a strange attractor in the coordinates$(A,\Psi )$.

The given regime forms form a limiting-cycle regime through an infinite sequence of period-doubling bifurcations.  ~\cite{ruelle71}.

Figure 2 shows the kinetics of simple periodic oscillations of the variables from a Krebs cycle. Figure 3 shows the kinetic behavior of the respiratory chain, of the kinetic membrane potential, of the ATP, and of molecular oxygen. Figure 4 shows the time evolution of the biosynthesis of the nitrogenase complex. We see from these figures that we have a self-regulation of the biosystem in an oscillatory regime: At high values of   $A$    $A$$A$, we have low levels of $E_{1} $, $E_{3} $, $E_{4} $;  low values of $A$ corresponds to high values of $E_{1} $, $E_{3} $, $E_{4} $. This situation determines the kinetic behavior of the other reactants.

At the parameter value $S_{0} =0.063$ the attractor is a limiting cycle. At the point $S_{0} =0.065$this limiting cycle undergoes a bifurcation into a limiting cycle with a period t twice the original period. The limiting cycle which is generated at $S_{0} =0.0689$again doubles through a period-doubling bifurcation. The next bifurcation point found is $S_{0} =0.06932$. After an infinite sequence of doubling bifurcations of the limiting cycle, an aperiodic oscillatory attractive regime - a strange-attractor regime - appears in the phase space.

\vspace*{7mm}

\begin{figure}[htp]
 \includegraphics[width=6cm]{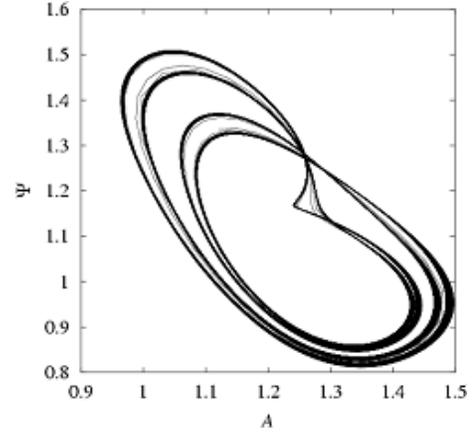}
 \caption{\label{fig1} Strange attractor in projection ($A, \Psi$).}
\end{figure}

\begin{figure}[htp]
 \includegraphics[width=8cm]{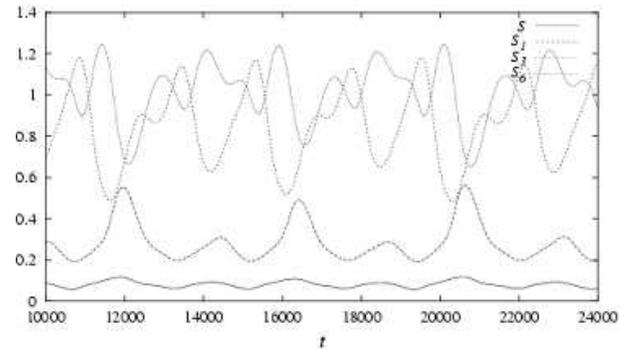}
 \caption{\label{fig2} Kinetics of the Krebs cycle.}
\end{figure}

\begin{figure}[htp]
 \includegraphics[width=8cm]{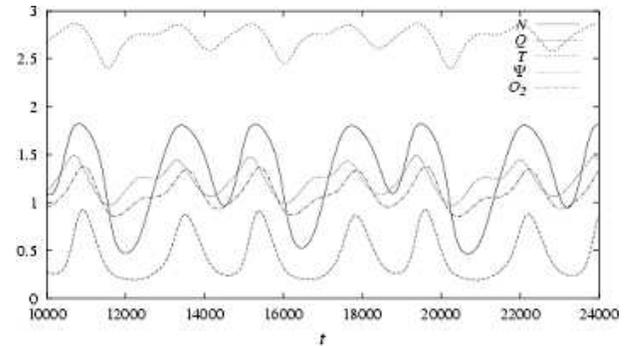}
 \caption{\label{fig3} Kinetics of the respiratory chain, of the ATP, of $O_2$, 
  and of the kinetic membrane potential.}
\end{figure}

\begin{figure}[htp]
	\includegraphics[width=8cm]{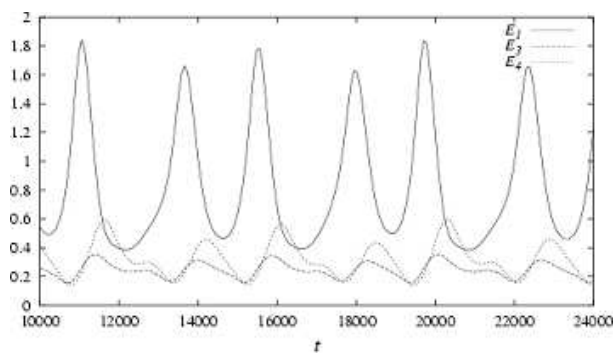}
	\caption{\label{fig4} Kinetics of the nitrogenase complex.}
\end{figure}
\bibliographystyle{apsrev}

\vspace*{7mm}

\end{document}